\documentclass[pra,aps,superscriptaddress,a4paper,preprint,endfloats]{revtex4-1}

\usepackage{graphicx}
\usepackage{float} % allows [H] to be used for placing figures and captions at the current location
\usepackage{color}
\usepackage{lineno}
\usepackage{hyperref}
\newcommand{\degree}{\ensuremath{^\circ}} %define the degree symbol

\begin{document}

\title{Tunable magnetoresistance in an asymmetrically coupled single molecule junction}

\thanks{\em{Nature Nanotechnol.} 10, 259 (2015); \url{http://dx.doi.org/10.1038/nnano.2014.326}}

\author{Ben Warner}
\affiliation{London Centre for Nanotechnology, University College London (UCL), London WC1H 0AH, U.K.}
\affiliation{Department of Physics \& Astronomy, UCL, London WC1E 6BT, U.K.}

\author{Fadi El Hallak}
\altaffiliation [Present address: ] {Seagate Technology, Derry BT48 0BF, U.K.}
\affiliation{London Centre for Nanotechnology, University College London (UCL), London WC1H 0AH, U.K.}

\author{ Henning Pr\"user}
\affiliation{London Centre for Nanotechnology, University College London (UCL), London WC1H 0AH, U.K.}

\author{John Sharp}
\affiliation{Surface Science Research Centre and Department of Chemistry, University of Liverpool, Liverpool, L69 3BX, U.K.}

\author{Mats Persson}
\affiliation{Surface Science Research Centre and Department of Chemistry, University of Liverpool, Liverpool, L69 3BX, U.K.}
\affiliation{Department of Applied Physics, Chalmers University of Technology, SE-412 96, G\"oteborg, Sweden}

\author{Andrew J. Fisher}
\affiliation{London Centre for Nanotechnology, University College London (UCL), London WC1H 0AH, U.K.}
\affiliation{Department of Physics \& Astronomy, UCL, London WC1E 6BT, U.K.}

\author{Cyrus F. Hirjibehedin}
\email[email: ]{c.hirjibehedin@ucl.ac.uk}
\affiliation{London Centre for Nanotechnology, University College London (UCL), London WC1H 0AH, U.K.}
\affiliation{Department of Physics \& Astronomy, UCL, London WC1E 6BT, U.K.}
\affiliation{Department of Chemistry, UCL, London WC1H 0AJ, U.K.}

\maketitle

\textbf{Phenomena that are highly sensitive to magnetic fields can be exploited in sensors and non-volatile memories \cite{Wolf:2001}. The scaling of such phenomena down to the single molecule level \cite{Schmaus:2011jc,Kawahara2012} may enable novel spintronic devices \cite{Bogani:2008hi}. Here we report magnetoresistance in a single molecule junction arising from negative differential resistance that shifts in a magnetic field at a rate two orders of magnitude larger than Zeeman shifts. This sensitivity to the magnetic field produces two voltage-tunable forms of magnetoresistance,  which can be selected via the applied bias. The negative differential resistance is caused by transient charging  \cite{Chen:1999ec,Wu:2004cq,FernandezTorrente:2012eza} of an iron phthalocyanine (FePc) molecule on a single layer of copper nitride (Cu$_2$N) on a Cu(001) surface, and occurs at voltages corresponding to the alignment of sharp resonances in the filled and empty molecular states with the Cu(001) Fermi energy. An asymmetric voltage-divider effect enhances the apparent voltage shift of the negative differential resistance with magnetic field, which inherently is on the scale of the Zeeman energy \cite{Ham:2012}. These results illustrate the impact that asymmetric coupling to metallic electrodes can have on transport through molecules, and highlight how this coupling can be used to develop molecular spintronic applications.}

Research into magnetoresistance \cite{Binasch:1989tj,Moodera:1995wo} has been driven by the widespread use of giant magnetoresistance (GMR) sensors in hard drives as well as other applications such as magnetoresistive random access memory (MRAM) \cite{Wolf:2001}.
%GMR sensors are constructed from two ferromagnetic layers separated by a thin non-magnetic layer; in these junctions, large resistance changes occur in the presence of a magnetic field $B$ \cite{Binasch:1989tj}. This change in resistance can be further increased by replacing the non-magnetic layer with an insulating layer: in this case the effect is called tunnelling magneotresistance (TMR) \cite{Moodera:1995wo}. 
To reach even higher storage densities, research has begun to concentrate on magnetoresistance at the atomic scale \cite{Schmaus:2011jc,Kawahara2012,Urdampilleta:2011ii}. For a single molecule, however, the small area for enclosing flux and modest energy scales associated with electronic Zeeman shifts typically make it difficult to tune magnetoresistive phenomena with an external magnetic field.

Another electron transport phenomenon with technological relevance is negative differential resistance (NDR) \cite{Lyo:1989vf, Chen:2007hc, Xue:372652, Wang:2003ek, Chen:1999ec, Gaudioso:2000p464, Grobis:2005bz,Tu:2008js,Heinrich:2011cc, FernandezTorrente:2012eza}, in which an increase in voltage causes a decrease in current. Commercial devices, such as the resonant tunnelling diode, utilise these regions in specialised applications \cite{Brown:1999wl,Ozbay:1991ti}. Various mechanisms cause NDR at the atomic scale \cite{Lyo:1989vf, Chen:2007hc, Xue:372652, Wang:2003ek, Chen:1999ec, Gaudioso:2000p464, Grobis:2005bz,Tu:2008js,Heinrich:2011cc, FernandezTorrente:2012eza}, though none are expected to have a magnetic field dependence that would shift the NDR on a scale larger than the Zeeman energy.

\begin{figure}[H]%[h!tbp]
\centering
\includegraphics[scale=1]{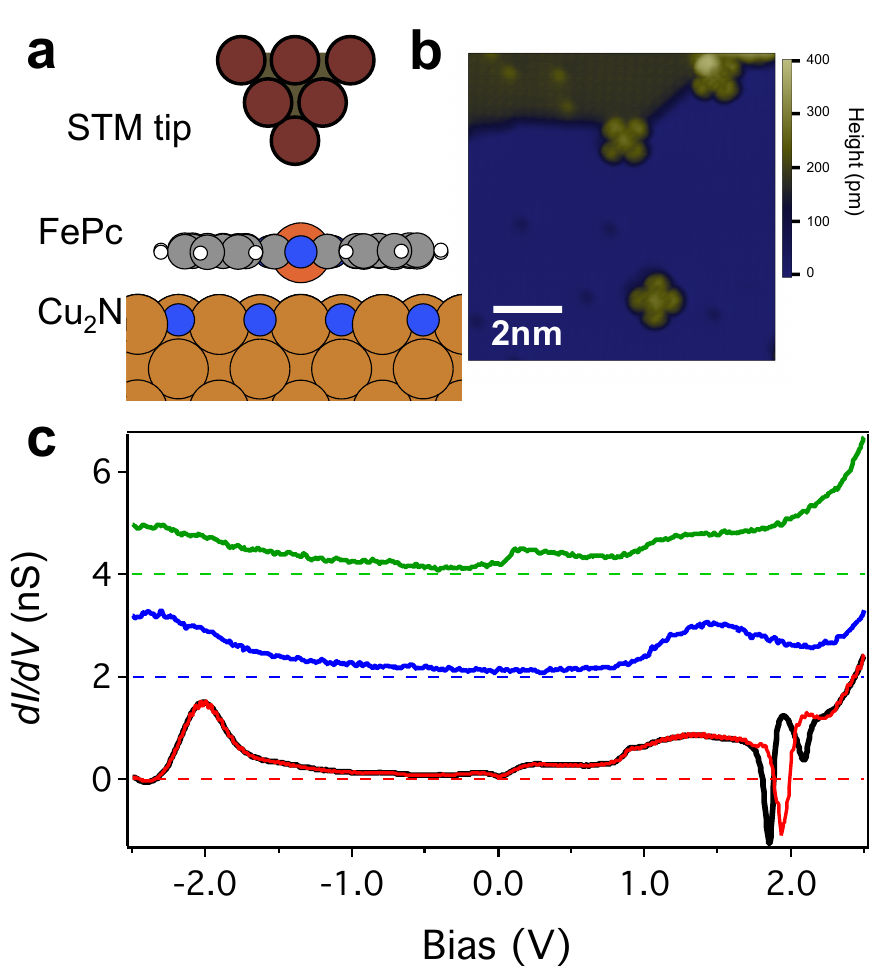}
\caption{FePc on \(\text{Cu}_2\text{N}\)/Cu(001). a) Schematic of the experimental configuration with the tunnel junction formed by the STM tip and \(\text{Cu}_2\text{N}\)/Cu with the FePc molecule sitting in between. Atoms in the molecule and Cu$_2$N/Cu are colour coded: grey=C, white=H, blue=N, orange=Fe, light brown=Cu. The tip (red and brown) is PtIr, however the last atoms are likely Cu because the tip is often indented into the surface to reshape it. b) STM topograph of the surface showing various FePc molecules (\(V_{set}=-1.0\text{ V}\), \(I_{set}=0.1\text{ nA}\)). c) $dI/dV$ spectroscopy measurements taken above the centre of different molecules (\(V_{set}=-2.5\text{ V}\), \(I_{set}=0.8\text{ nA}\)). Two general classes are observed depending on binding site. Representative spectra for Cu and N sites (green and blue respectively) taken at $B=6\text{ T}$ are shown, but the features can vary significantly from molecule to molecule. Red spectra show a clear NDR feature, which appears in \(12.5\%\) of the molecules. This can shift by up to -15 mV/T, as seen in spectra taken at 0 T (red) and 6 T (black). The magnetic field only moves features in the NDR region: other features in the spectrum remain constant. Traces have been offset vertically for clarity; $dI/dV=0$ is indicated by a dashed line for each trace.}
\label{fig:Overview}
\end{figure}

Using low temperature scanning tunnelling microscopy (STM) (see Supplementary Methods), we observe an NDR effect for FePc molecules placed in a vacuum junction on top of a Cu(001) surface capped with a single layer of Cu$_2$N (Fig. \ref{fig:Overview}). Cu$_2$N is a thin insulator that can decouple the spins of magnetic atoms from the underlying surface \cite{Hirjibehedin:2006p4}; FePc is a magnetic molecule that can be easily sublimed \cite{Scarfato:2008fza, Mugarza:2012bu, Minamitani:2012jua} and is observed to have interesting magnetic properties on thin insulating layers \cite{Tsukahara:2009p428}.
%FePc is a magnetic molecule that can be easily sublimed in UHV \cite{Scarfato:2008fza, Mugarza:2012bu, Minamitani:2012jua}, and is observed to have interesting magnetic properties on thin insulating layers \cite{Tsukahara:2009p428}.%
On Cu$_2$N, FePc is centred above both Cu and N sites. The two binding sites can be differentiated using atomically resolved imaging and spectroscopic measurements; typical spectra of both types are seen in Fig. \ref{fig:Overview}c. Furthermore, a broad distribution of binding angles is observed, with shallow peaks at $0\degree$, $18\degree$, and $45\degree$ with respect to the crystallographic axes. Density functional theory (DFT) calculations (see Supplementary Methods) indicate relatively weak variations in the binding energy with angle (see Supplementary Table and Supplementary Data).

Remarkably, when a magnetic field is applied the NDR minimum can shift by two orders of magnitude more than the electronic Zeeman effect (Fig. \ref{fig:Overview}c), here almost 0.1 V for an applied field of 6 T. To our knowledge such magnetic sensitivity has not been observed for other systems exhibiting NDR. The NDR effect is observed in \(12.5\%\) of the molecules (23 out of 184), at both positive and negative bias, and at a variety of voltages for different molecules on both Cu and N sites (Supplementary Fig. 2). NDR is observed at various binding angles for both sites, suggesting that NDR occurs on molecules with different binding geometries. In almost all cases (see Supplementary Methods), NDR was observed only at the centre of the FePc molecule (i.e. above the Fe atom, see Supplementary Fig. 3).
%We do not observe a difference in topography between molecules that show NDR and those that do not.%

\begin{figure}[H]%[h!tbp]
\centering
\includegraphics[scale=1]{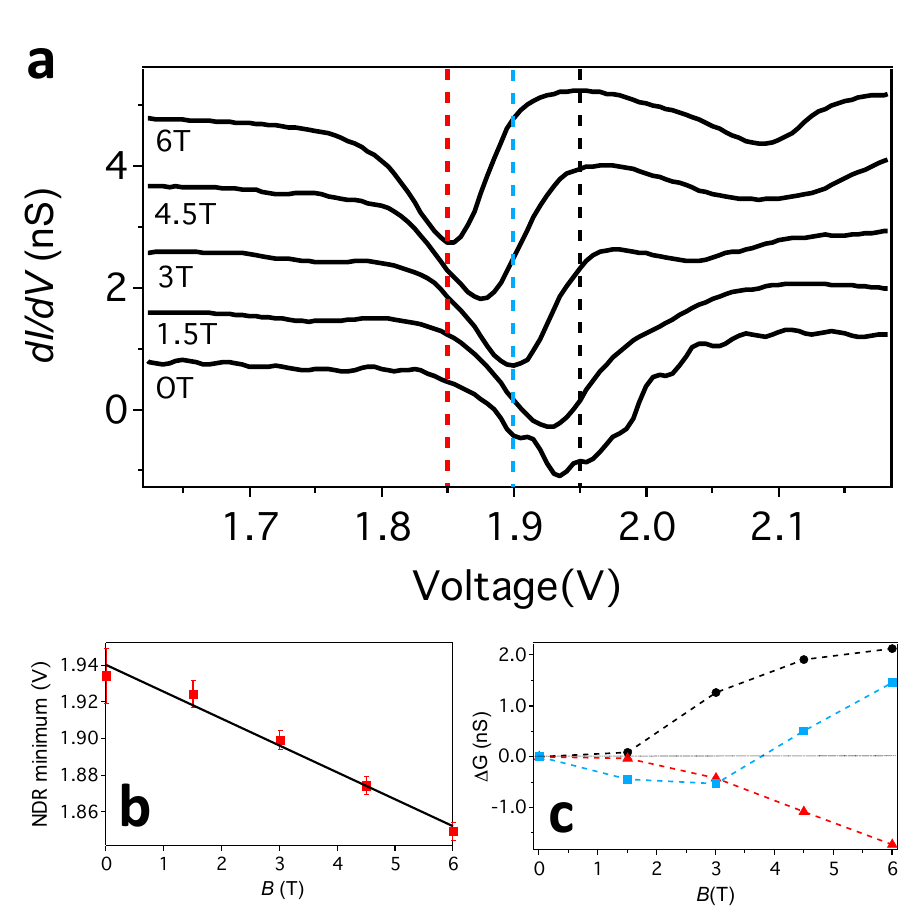}
\caption{Differential conductance changes caused by magnetic field sensitive NDR. a) Differential conductance spectra (\(V_{set}=-2.5\text{ V}\), \(I_{set}=0.8\text{ nA}\)) acquired above the centre of an FePc molecule displaying NDR at $B=0$ T, 1.5 T, 3.0 T, 4.5 T, and 6.0 T (as labelled). As $B$ is increased the NDR region moves to lower voltages. Spectra are offset vertically for clarity. Vertical dashed lines indicate 1.85 V (red), 1.90 V (blue), and 1.95 V (black). b) NDR minimum vs. $B$, with the solid line showing a gradient of -15mV/T. Error bars show the uncertainty in defining the minimum for each spectrum. c)  \(\Delta G(V,B)=G(\text{V},B)-G(\text{V},0)\) versus $B$ at 1.95 V (black), 1.9 V (blue), and 1.85 V (red).}
\label{fig:FieldandXMR}
\end{figure}

A more detailed dependence of the changes in the NDR spectra with perpendicular magnetic field is shown in Fig. \ref{fig:FieldandXMR}a. As seen in Fig. \ref{fig:FieldandXMR}b, the voltage of the NDR minimum shifts approximately linearly with a slope of -15 mV/T. An increase in $|B|$ shifts the NDR minimum to lower (less positive / more negative) voltages, but the slope varies from molecule to molecule, ranging from -2 mV/T up to -15 mV/T. Furthermore, our measurements suggest that the shift of the NDR depends only on the magnitude of the field component perpendicular to the plane, with an in-plane field of 1 T and a reversal of the sign of the magnetic field having no impact. 
%We note for completeness that in Fig. \ref{fig:FieldandXMR}a a second decrease in $dI/dV$ develops at larger voltages and at higher magnetic fields. It may be a vibronic copy \cite{Wu:2004cq} of the original NDR feature or a second, independent NDR feature; however, it is difficult to assign a clear physical origin to it because it does not occur on most molecules.% 
Additionally, on rare occasions we have observed sharp peaks in the conductance spectra in similar voltage ranges that exhibit a similar dependence on $B$ (see Supplementary Fig. 4).

\begin{figure}[H]%[h!tbp]
\centering
\includegraphics[scale=1]{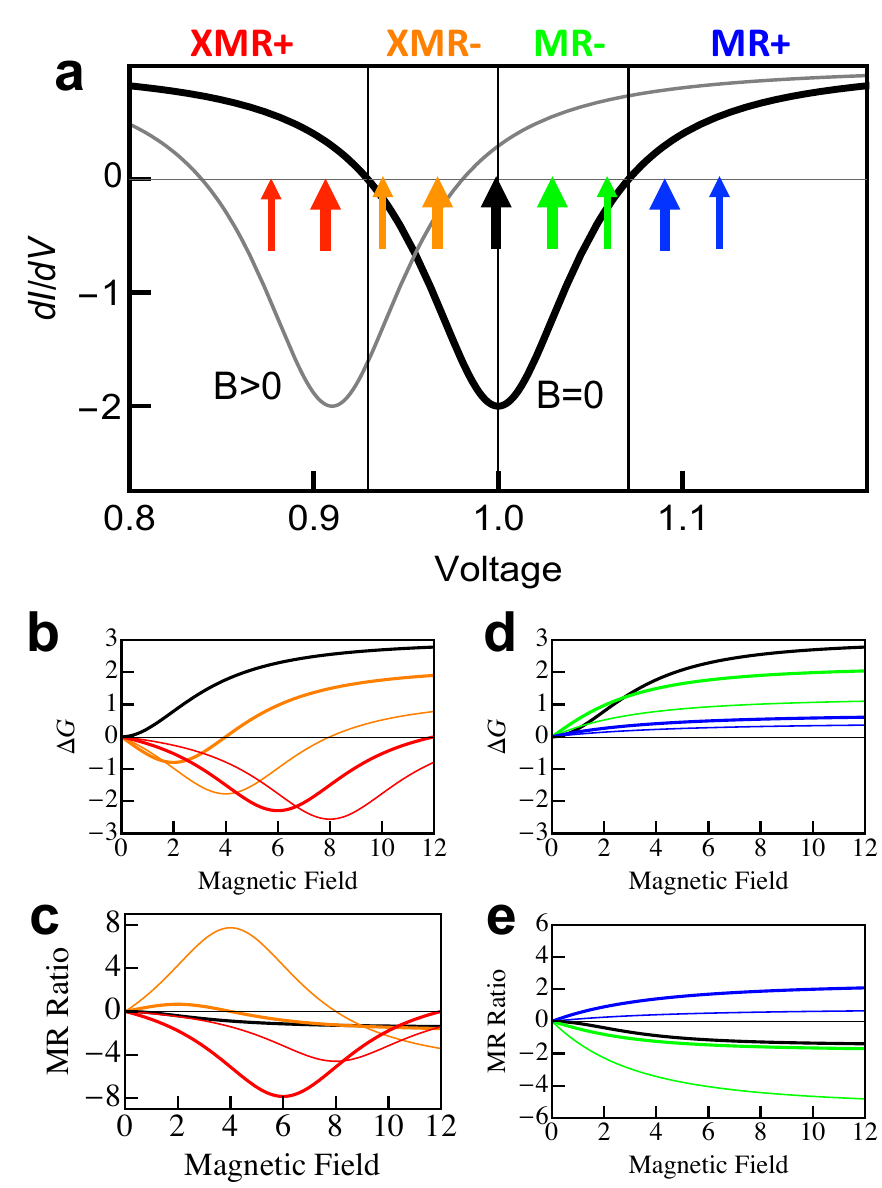}
\caption{Model of magnetically sensitive NDR. a) Differential conductance versus voltage for constant differential conductance background with an NDR feature that shifts linearly with $|B|$; two values of $B$, $B=0$ (thick black curve) and $|B|>0$ (thin grey curve), are shown. Arrows mark voltages shown in panels below. The voltage ranges in which there is a monotonic and ``cross-over magnetoresistance" change in $\Delta G$ and the magnetoresistance ratio are labelled MR$\pm$ and XMR$\pm$ respectively. All units have arbitrary dimensions. b)  $\Delta G$ vs. $B$ in XMR+ and XMR- regime. c) Corresponding magnetoresistance ratio from panel b. d)  $\Delta G$ vs. $B$ in MR+ and MR- regime. e) Corresponding magnetoresistance ratio from panel d.}
\label{fig:MRmodel}
\end{figure}

The ability to manipulate NDR with a magnetic field not only enables tuning of the voltage of the NDR minimum \cite{Mizuta:1995vc} but also results in the creation of a junction that exhibits two novel magnetoresistance effects. Figure \ref{fig:MRmodel} shows a model of the NDR where the differential conductance line shape $G(V,B)$ is represented by a Lorentzian dip that shifts linearly with $|B|$ on top of a constant background. For voltages that are more positive than the voltage of the NDR minimum, the change in differential conductance \(\Delta G(V,B)=G(\text{V},B)-G(\text{V},0)\) is always positive and increases with $|B|$ until it saturates. Remarkably, however, the magnetoresistance ratio $\frac{\Delta G (V,B)}{G(V,0)}$ can become arbitrarily large as $V$ approaches the value at which $G(V,0) = 0$. Furthermore, its sign is positive or negative depending on the sign of $G (V,0)$; we therefore label these the MR+ and MR- regions respectively. In practice, of course, the arbitrarily large magnetoresistance ratio that occurs near the boundary of these two regions would be limited by experimental constraints. The second magnetoresistance effect, which we refer to as ``cross-over magnetoresistance" (XMR), is manifested at voltages that are more negative than the NDR minimum. In this regime, as seen in Fig. \ref{fig:MRmodel}, $\Delta G(V,B)$ initially becomes increasingly negative with $|B|$ until it reaches a minimum value; after this, it becomes more positive, crosses zero, and then saturates at a limiting value. The magnetic field at which the polarity of the differential conductance ``crosses over" varies with voltage, creating a magnetic-field sensitive switch that is tunable with voltage. As seen in Fig. \ref{fig:FieldandXMR}c, both of these effects are observed for FePc on Cu$_2$N.

\begin{figure}[H]%[h!tbp]
\centering
\includegraphics[scale=1]{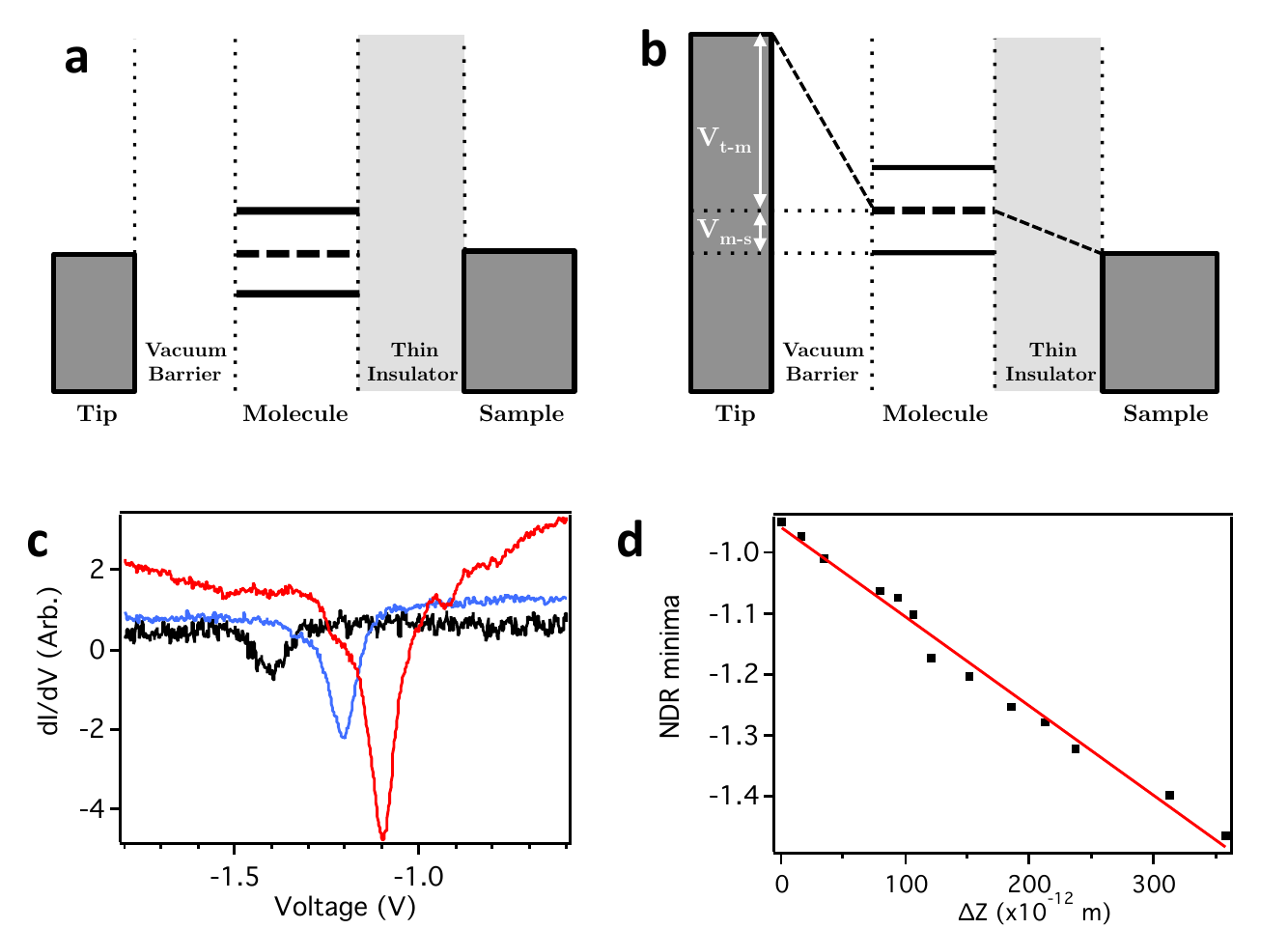}
\caption{Tunnelling across a double barrier junction. a) Junction at $V=0$, with the Fermi levels of the tip and substrate aligned. Solid horizontal lines indicate the filled and empty states of the molecule; dashed horizontal line is a reference between these levels. b) Junction at $V=V_{res}>0$, when the voltage dropped across the thin insulator (i.e. between the molecule and the underlying Cu surface) $V_{m-s}$ aligns the molecular orbital with the Fermi level of the substrate. The remaining voltage is dropped across the vacuum barrier (i.e. between the tip and the molecule) $V_{t-m}$, allowing the molecular levels to shift with respect to the substrate Fermi level. Note that the potentials account for the negative sign of the charge carriers. c) Selected $dI/dV$ vs. voltage spectra ($V_{set}=-1.8 \text{ V}$) obtained at $I_{set}=$25 pA (black), 250 pA (blue), and 500 pA (red). d) NDR minimum versus change in tip-substrate distance $\Delta z$, which is calibrated using $I(z)$ spectroscopy. As the tip moves towards the substrate (higher current setpoint), the NDR minimum shifts linearly closer to the Fermi energy, as highlighted by the solid red line.
 }
\label{fig:DBmodel}
\end{figure}

To explain this novel manifestation of magnetically sensitive NDR (Supplementary Discussion), we suggest a mechanism based on transient charging that arises from the occupation of molecular resonances \cite{Mikaelian:2006ke}. This results in a change in the tunnelling rates through the molecule that can increase or decrease the differential conductance, with the latter resulting in NDR. Sharp states corresponding to a two-step resonant tunnelling process between the tip, the molecule, and the substrate have been observed in studies of individual molecules on thin insulators \cite{Wu:2004cq}. In the resonant tunnelling process, voltage is dropped across both barriers (vacuum and Cu$_2$N) in the tunnel junction, with most of the drop expected to occur in the vacuum between the tip and the molecule (Fig. \ref{fig:DBmodel}b).  The small fraction of the applied bias voltage dropped across the thin insulator therefore shifts the molecular orbitals with respect to the substrate Fermi energy.   

Because the fraction of the voltage dropped across the thin insulator varies with the relative size of the tip-molecule gap, the hallmark of this mechanism is a shifting of the NDR minimum with the height of the tip above the surface \cite{Wu:2004cq}.
%This behaviour does not occur for local density of states features observed in single-step tunnelling.
As seen in Fig. \ref{fig:DBmodel}c, the NDR minimum clearly shifts closer to the Fermi energy as the set point current (tip height) is increased (decreased). Fig. \ref{fig:DBmodel}d further shows that position of the NDR minimum shifts linearly with tip height, as expected. DFT calculations in which an electric field has been added to the system also show that a finite potential drop exists between the molecule and the substrate.
 
In these asymmetric, double-barrier tunnel junctions, the strongest resonance occurs when one of the molecular orbitals aligns with the Fermi energy in the substrate (Fig. \ref{fig:DBmodel}b) because the molecule is more strongly coupled to the substrate than to the tip \cite{Mikaelian:2006ke}. Depending on whether the alignment occurs with an empty or filled orbital, the molecule can be transiently negatively or positively charged respectively during the transport process; since these occur at negative and positive bias respectively \cite{Wu:2004cq} and can result in either increased or decreased differential conductance \cite{FernandezTorrente:2012eza}, the NDR can occur in either polarity of bias voltage for different molecules. Note that this charging is a consequence of the extended lifetime of the tunnelling electron on the molecule: if the tip were moved away the molecule would quickly return to its neutral state.

\begin{figure}[H]%[h!tbp]
\centering
\includegraphics[scale=1]{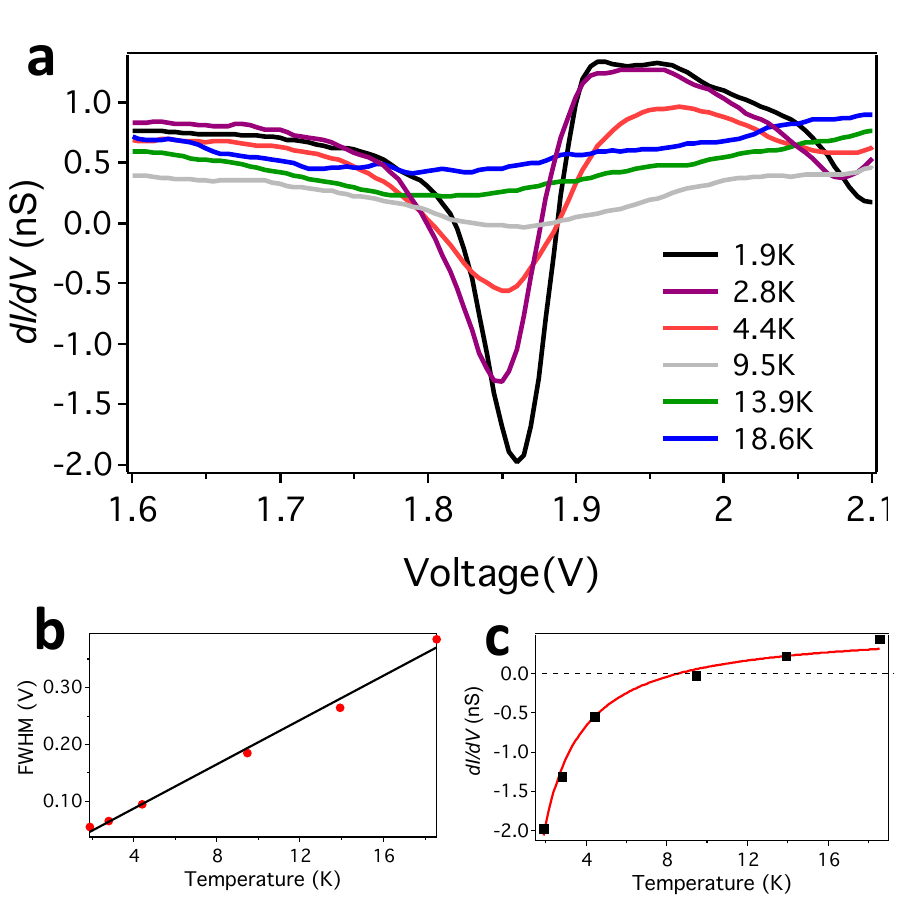}
\caption{Temperature dependance of NDR. a) Zoom-in of the NDR minimum in differential conductance (\(\text{V}_{set}=-2.5\text{ V}\), \(\text{I}_{set}=0.8\text{ nA}\)) at different temperatures; the minimum  remains positive above approximately 10 K.  b) FWHM of the NDR minimum (with the baseline  taken as the value at 1.6 V) vs. temperature. Black line highlights the linear trend, with a slope of $225 \text{ } k_BT/e$. c) Differential conductance at the NDR minimum vs. temperature. Solid red line is a guide to the eye for a $1/T$ decay.}
\label{fig:Temp}
\end{figure}

Because the molecular levels shift with respect to the Fermi energy by much less than the applied bias voltage, the apparent voltage scale of the resonance is enhanced  \cite{Wu:2004cq,FernandezTorrente:2012eza}. This can be quantified by considering the behaviour with temperature. As seen in Fig. \ref{fig:Temp}a, the NDR minimum becomes dramatically more shallow and broad with increasing temperature. Figure \ref{fig:Temp}c shows that the depth of the NDR minimum decreases with a $1/T$ dependence, where $T$ is the substrate temperature, as expected for thermal smearing. The full width at half maximum (FWHM) is shown in Fig. \ref{fig:Temp}b and is found to increase linearly with a rate of approximately $(225 \pm 11) \text{ } k_B/e$, where $k_B$ is the Boltzmann constant and $e$ is the magnitude of the electron charge. The expected broadening for thermal smearing from the Fermi seas in the tip and the substrate is  3.5 $k_{B}/e$, so for this molecule the enhancement is $225/3.5 \sim 65$.

Owing to the enhancement factor, the levels responsible for the NDR minimum shift with magnetic field at a much smaller intrinsic rate than the observed movement of the mimima: for the spectra shown in Fig. \ref{fig:FieldandXMR}, this would correspond to an intrinsic shift of $231 \text{ } \mu \text{eV/T}$, which is of the order of the Zeeman energy. Zeeman splitting of such sharp molecular resonances into doublets has been observed in the presence of a magnetic field \cite{Ham:2012}. This shows that an asymmetric junction not only can significantly influence the electronic properties of the junction \cite{Krull:2011ewa} but also allows for the enhancement of energy scales, causing small shifts in energy to be magnified. %to large scale changes.

In this case, the fact that NDR is observed only over the centre of FePc molecules suggests that the resonant levels are associated with the Fe d-orbitals. Furthermore, because of the large exchange splitting between the majority and minority Fe d-levels (see Supplementary Fig. 1), the resonant levels are spin-polarised and non-degenerate; they would therefore shift in the presence of a magnetic field rather than splitting. Since most of the levels close to the Fermi energy are minority spin states (see Supplementary Fig. 1), it is sensible that we have only observed resonances shifting in one direction with field. The lack of an observed shift with the application of a small in-plane magnetic field is consistent with an axial anisotropy for the total d-electron moment, oriented out of the plane, as has been observed for FePc on CuO \cite{Tsukahara:2009p428}. 
%DFT ground state calculations cannot be used directly to describe electron attachment and detachment levels involved in the proposed transient charging process. However, the proximity of the minority levels with d$_\pi$ and d$_{z^2}$ characters to the Fermi energy when the molecule is adsorbed on the surface, as well as the orientation of these levels out of the plane, which allows them to overlap with the tip, suggests that these are the ones that are most likely to be involved in this process.%

In a simple parallel-plate capacitor model of the tunnel junction formed by the tip, the molecule, and the underlying metal, the enhancement factor can also be described by the fraction of the voltage that is dropped across the Cu$_2$N: $d^*/(d^*+z)$, where $d=d^* \epsilon$ is the distance between the molecule and the underlying metal, $\epsilon$ is the effective dielectric constant of the Cu$_2$N monolayer, and $z$ is the distance between the molecule and the tip. Although there are no existing direct measurements of $\epsilon$ for Cu$_2$N, we can estimate $d \sim (0.55 \pm 0.05 )\text{ nm}$ and $z \sim (0.60 \pm 0.10) \text{ nm}$ (Supplementary Methods) to obtain $\epsilon \sim (60 \pm 12)$, which is approximately an order of magnitude greater than for other thin insulators like aluminium oxide. Therefore, using the simple model shown here, both the high effective dielectric constant and the thinness of the Cu$_2$N play a role in creating an enhancement value that is much higher than that observed on other thin insulators \cite{Wu:2004cq}.

Furthermore, the large enhancement factor explains the low number of molecules for which NDR is observed. In principle, all of the molecules should exhibit this phenomenon.
%if sufficient voltage is applied to bring the level that is closest to the Fermi energy into resonance.
However, the spectroscopic window in which we can measure is limited to $\sim \pm 2.5 \text{ V}$ by the stability of the molecules. As the enhancement factor is $\sim60$, this results in our measurements only being sensitive to levels within $\sim 40 \text{ mV}$ of the Fermi energy.  Because the spectroscopy on each molecule varies, we can only observe NDR for molecules in which the appropriate levels lie close enough to the Fermi energy to fall within our measurement window.

In summary, we observe magnetically sensitive NDR in a single-molecule junction arising from resonant tunnelling producing charging in the molecule. The effective shift of the NDR with magnetic field is enhanced by the inherent voltage division across the two asymmetric tunnelling barriers; this allows for the creation of novel magnetoresistance phenomena. Similar enhancement of the effective energy scale for other multi-step tunnelling phenomena, both magnetic and non-magnetic in origin, should be possible. Furthermore, the size of the enhancement can be controlled by tuning the asymmetry of the tunnelling barriers, which can be modified by making physical or chemical changes to the junction by using different thin insulators or molecules \cite{FernandezTorrente:2012eza,Kahle:2012fm,Repp:2005hc,Tsukahara:2009p428,Chen:2008p258} . This highlights the prominent role that the junction itself can play in defining the properties of the smallest possible electronic and spintronic device architectures.

%%References

\section*{Acknowledgements}
We acknowledge Gabriel Aeppli, Vincent Crespi, Jeroen Elzerman, Joaqu\'in Fern\'andez-Rossier, Mark Hybertsen, Peter Littlewood, Sebastian Loth, Christoph Mathieu, Markus Ternes, and Joris van Slageren for simulating discussions. B.W., F.E.H., H.P., A.J.F., and C.F.H. acknowledge financial support from the EPSRC [EP/H002367/1 and EP/D063604/1] and the Leverhulme Trust [RPG-2012-754]. M.P. is grateful for support from the EU project ARTIST and allocations of computer resources at HECToR through the Materials Chemistry Consortium funded by EPSRC [EP/L000202/1] and at PDC through SNIC.

\section*{Author contributions}
F.E.H. and C.F.H. conceived of the experiments; B.W., F.E.H. and H.P. performed the experiments and analysed the results; J.S. and M.P. performed the DFT calculations; all authors discussed the results and contributed to the writing of the paper.

\section*{Additional information}
Supplementary information accompanies this paper at www.nature.com/naturenanotechnology. Reprints and permission information is available online at http://npg.nature.com/reprintsandpermissions/. Correspondence and requests for materials should be addressed to C.F.H.

\section*{Competing financial interests}
The authors declare no competing financial interests.

% --------------------
% Figure 1
% --------------------

\newpage

\newpage

% --------------------
% Figure 2
% --------------------

\newpage

% --------------------
% Figure 3
% --------------------

\newpage

% --------------------
% Figure 4
% --------------------

% --------------------
% Figure 5
% --------------------

\newpage

\end{document}